\documentclass[twocolumn]{aastex631}

\newcommand{\rvone}{\ensuremath{RV_1}}
\newcommand{\rvtwo}{\ensuremath{RV_2}}

\newcommand{\project}[1]{\textsl{#1}} 
\newcommand{\package}[1]{\textsl{#1}}

\newcommand{\thejoker}{\project{The~Joker}}
\newcommand{\github}{\package{GitHub}}
\newcommand{\python}{\package{Python}}
\newcommand{\lamost}{LAMOST}
\newcommand{\lamostmrs}{LAMOST-MRS }
\newcommand{\gaia}{\textsl{Gaia}}

\newcommand{\apogee}{APOGEE}

\newcommand{\kms}{\ensuremath{\mathrm{km}~\mathrm{s}^{-1}}}

\newcommand{\dayd}{\ensuremath{\mathrm{d}}}

\newcommand{\stable}{Table}
\usepackage{CJK}
\usepackage{amsmath}
\usepackage{booktabs,tabularx}
\usepackage{array}
\usepackage{ragged2e}
\usepackage{microtype}
\usepackage{amsfonts}
\usepackage{amssymb}
\usepackage{graphicx}
\usepackage{color}
\usepackage{threeparttable}
\usepackage{soul} 
\usepackage{xcolor} 
\usepackage{enumitem}
\usepackage{subfigure}

\begin{document}
\title{Orbital Parameters of 665 Double-lined Spectroscopic Binaries in the \lamost\ Medium-Resolution Survey}

\correspondingauthor{Jiao Li}
\email{lijiao@ynao.ac.cn}
\correspondingauthor{Zhanwen Han}
\email{zhanwenhan@ynao.ac.cn}

\author[0009-0008-0809-8694]{Sufen Guo}
\affiliation{School of Physical Science and Technology, Xinjiang University, Urumqi 830046, People's Republic of China}
\affiliation{Yunnan Observatories, Chinese Academy of Sciences (CAS), Kunming 650216, People's Republic of China}
\affiliation{Key Laboratory for the Structure and Evolution of Celestial Objects, CAS, Kunming 650216, People's Republic of China}
\affiliation{International Centre of Supernovae, Yunnan Key Laboratory, Kunming 650216, People's Republic of China}

\author[0000-0002-9975-7833]{Mikhail Kovalev} %(Михаил~Ковалёв)
\affiliation{Yunnan Observatories, Chinese Academy of Sciences (CAS), Kunming 650216, People's Republic of China}
\affiliation{Key Laboratory for the Structure and Evolution of Celestial Objects, CAS, Kunming 650216, People's Republic of China}
\affiliation{International Centre of Supernovae, Yunnan Key Laboratory, Kunming 650216, People's Republic of China}

\author[0000-0002-2577-1990]{Jiao Li}
\affiliation{Yunnan Observatories, Chinese Academy of Sciences (CAS), Kunming 650216, People's Republic of China}
\affiliation{Key Laboratory for the Structure and Evolution of Celestial Objects, CAS, Kunming 650216, People's Republic of China}
\affiliation{International Centre of Supernovae, Yunnan Key Laboratory, Kunming 650216, People's Republic of China}

\author[0000-0002-3839-4864]{Guoliang L\"u}
\affiliation{School of Physical Science and Technology, Xinjiang University, Urumqi 830046, People's Republic of China}

\author[0000-0001-6808-638X]{Shi Jia}
\affiliation{Yunnan Observatories, Chinese Academy of Sciences (CAS), Kunming 650216, People's Republic of China}
\affiliation{Key Laboratory for the Structure and Evolution of Celestial Objects, CAS, Kunming 650216, People's Republic of China}
\affiliation{International Centre of Supernovae, Yunnan Key Laboratory, Kunming 650216, People's Republic of China}

\author[0000-0002-1421-4427]{Zhenwei Li}
\affiliation{Yunnan Observatories, Chinese Academy of Sciences (CAS), Kunming 650216, People's Republic of China}
\affiliation{Key Laboratory for the Structure and Evolution of Celestial Objects, CAS, Kunming 650216, People's Republic of China}
\affiliation{International Centre of Supernovae, Yunnan Key Laboratory, Kunming 650216, People's Republic of China}

\author[0000-0003-3832-8864]{Jiangdan Li}
\affiliation{Yunnan Observatories, Chinese Academy of Sciences (CAS), Kunming 650216, People's Republic of China}
\affiliation{Key Laboratory for the Structure and Evolution of Celestial Objects, CAS, Kunming 650216, People's Republic of China}
\affiliation{International Centre of Supernovae, Yunnan Key Laboratory, Kunming 650216, People's Republic of China}

\author[0000-0003-4829-6245]{Jianping Xiong}
\affiliation{Yunnan Observatories, Chinese Academy of Sciences (CAS), Kunming 650216, People's Republic of China}
\affiliation{Key Laboratory for the Structure and Evolution of Celestial Objects, CAS, Kunming 650216, People's Republic of China}
\affiliation{International Centre of Supernovae, Yunnan Key Laboratory, Kunming 650216, People's Republic of China}

\author[0009-0000-6595-2537]{Mingkuan Yang}
\affiliation{Key Laboratory of Optical Astronomy, National Astronomical Observatories, Chinese Academy of Sciences, Beijing 100101, China}

\author[0009-0004-9758-0722]{Tongyu He}
\affiliation{College of Physics Science and Technology, Hebei University, Baoding 071002, China}
\affiliation{International Centre of Supernovae, Yunnan Key Laboratory, Kunming 650216, People's Republic of China}

\author[0000-0001-5284-8001]{Xuefei Chen}
\affiliation{Yunnan Observatories, Chinese Academy of Sciences (CAS), Kunming 650216, People's Republic of China}
\affiliation{Key Laboratory for the Structure and Evolution of Celestial Objects, CAS, Kunming 650216, People's Republic of China}
\affiliation{International Centre of Supernovae, Yunnan Key Laboratory, Kunming 650216, People's Republic of China}

\author[0000-0001-9204-7778]{Zhanwen Han}
\affiliation{Yunnan Observatories, Chinese Academy of Sciences (CAS), Kunming 650216, People's Republic of China}
\affiliation{Key Laboratory for the Structure and Evolution of Celestial Objects, CAS, Kunming 650216, People's Republic of China}
\affiliation{International Centre of Supernovae, Yunnan Key Laboratory, Kunming 650216, People's Republic of China}

\begin{abstract}
The period, mass ratio, eccentricity, and other orbital parameters are fundamental for investigating binary star evolution. However, the number of binaries with known orbital parameters remains limited. Utilizing the \lamostmrs\ survey, we derived orbital solutions for 665 SB2 binaries by fitting the radial velocities of 1119 SB2 systems with at least six observations, employing a modified version of \thejoker\ optimized for SB2 binaries. To ensure the reliability of the results, four selection criteria were applied: reduced chi-square, normalized mean absolute error, maximum phase gap, and RV distribution metric. After applying these criteria, 665 reliable orbits were retained. Comparison with Kepler, TESS, and ZTF light curve data shows excellent agreement, with discrepancies in some cases attributed to shorter pulsation periods observed in light curves. Additionally, good consistency is found between our periods and those of SB1 systems in Gaia data. These orbital solutions contribute to understanding binary star evolution and the statistical properties of binary populations.

\end{abstract}


\section{Introduction} 
\label{sec:intro}

Nearly half of the stars in the Milky Way are binaries, and binary systems are essential for addressing many fundamental astrophysical problems \citep{moe2017mind,duchene2013stellar}.
Binary interactions significantly influence stellar evolution, producing phenomena and outcomes that are impossible for isolated stars \citep{jones2017binary}. Through these interactions, binaries can give rise to Type Ia supernovae, double neutron stars, double black holes \citep{belczynski2002comprehensive,webbink1998gravitational}, millisecond pulsars \citep{deng2020formation}, X-ray binaries \citep{deng2021evolution}, and gravitational wave sources \citep{li2024compact}. The mass, orbital period, mass ratio, and eccentricity of binary stars are key parameters that investigate their evolution and interactions and these parameters are crucial for binary population synthesis models, which help us understand star formation and binary evolution \citep{han2020binary}. This highlights the need for accurate identification of binary systems and precise determination of their orbital parameters.

Among binary systems, double-lined spectroscopic binaries (SB2s) are particularly valuable for their ability to provide precise orbital parameters especially mass ratios because SB2s capture the signatures of both stars in their spectra. Unlike single-lined spectroscopic binaries (SB1s), which require monitoring over multiple epochs to detect periodic changes in radial velocity (RV), SB2s can often be identified from a single epoch due to the distinct RV components of both stars. However, deriving reliable orbital solutions for SB2s demands sufficient multi-epoch observations to fully constrain their orbital parameters.

With the advent of large-scale spectroscopic surveys like Large Sky Area Multi-Object Fiber Spectroscopic Telescope (\lamost, \citet{cui2012large, zhao2012lamost, luo2012data, wang1996special,su2004active}) , Galactic Archaeology with HERMES (GALAH, \citet{galah2015}), and the Apache Point Observatory Galactic Evolution Experiment (\apogee, \citet{apogee2017AJ_154_94M}), unprecedented access to vast stellar samples has significantly advanced efforts to identify SB2 systems. For instance, \citet{li2021double} identified 3133 SB2s using \lamost\ Medium-Resolution Spectroscopic Survey (\lamostmrs) data, while \citet{zhang2022spectroscopic} employed a neural network model to detect 2198 SB2 candidates from the same dataset. \citet{wang2021lamost} reported 2700 SB2 candidates from \lamost's time-domain survey of four K2 plates, and \citet{zheng2023searching} discovered 281 new SB2s using \lamostmrs. Additionally, \citet{liu2024double} identified 4848 SB2s through LAMOST-LRS data. From the RAVE survey, \citet{matijevivc2010double} found 123 SB2s, while \citet{kounkel2021double} identified 7273 SB2s from APOGEE. Furthermore, \citet{el2018discovery} detected 2500 SB2s using APOGEE data, and \citet{kovalev2022detection,kovalev2024detection} reported the discovery of 1410 and 8105 new SB2s, respectively, from \lamostmrs. Although these studies highlight the growing capability of various surveys to identify SB2 systems, precise determination of their orbital parameters are still limited. In this work, we use multi-epoch \lamostmrs\ observations to identify SB2s and derive their orbital parameters through radial velocity fitting.

The rest of this paper is organized as follows: In Section \ref{sec:data}, we introduce the \lamostmrs  spectral dataset used in this study. Section \ref{sec:method} outlines the method for radial velocity fitting. In Section \ref{sec:fit_quality} we introduce the quality of the fitting. The results are presented in Section \ref{sec:results}, followed by a discussion of their implications in Section \ref{sec:Discussion}. Finally, we summarize our findings in Section \ref{sec:conclusion}.

\section{Data} \label{sec:data}

\begin{figure*}
    \includegraphics[width=\textwidth]{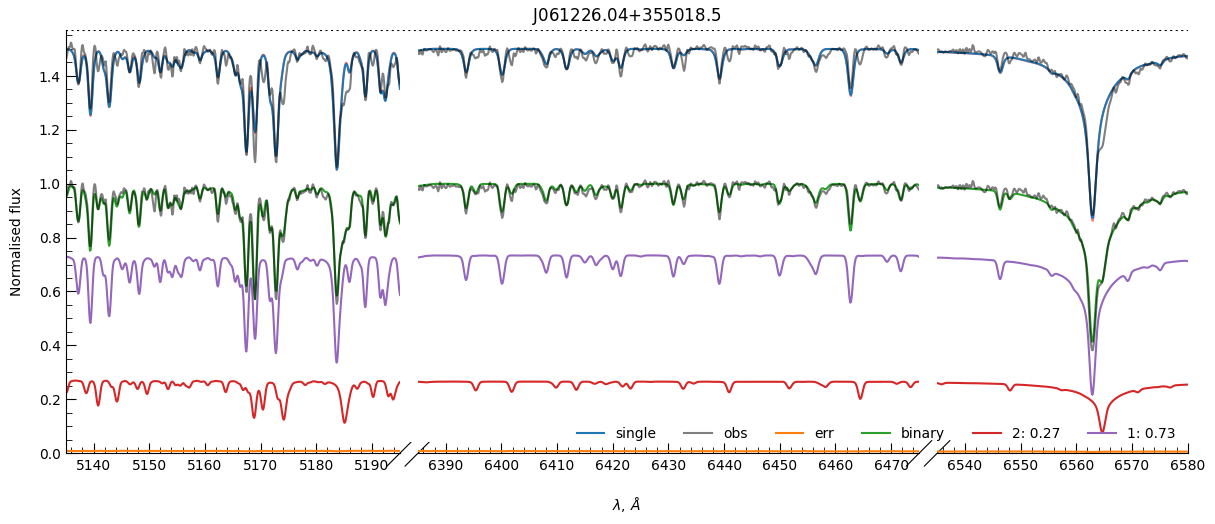}
    \caption{Example of the spectral fitting by the binary model (green line) and single-star model (blue-line with an offset of 0.5) for J061226.04+355018.5. Observed spectrum and its error are shown as a grey and orange lines. Primary and secondary are shown as magenta and red lines respectively.}
    \label{fig:examplefit}
\end{figure*}

The \lamost\ is a quasi-meridian reflective Schmidt telescope with an effective aperture of 4 meters, a 5-degree field of view, and a focal plane with a diameter of 1.75 meters. Equipped with 4,000 optical fibers on its focal plane, it can simultaneously acquire spectra for at most 4,000 celestial objects \citep{cui2012large, zhao2012lamost, luo2012data, wang1996special,su2004active}, making it one of the most efficient telescope worldwide for spectral acquisition. For our study, we utilized its Medium Resolution Survey (\lamostmrs) data, which offers a resolution of $R=\lambda/\Delta\lambda \sim 7500$ \citep{liu2020lamost,li2023lamost}. We used the binary spectral model by \citet{kovalev2022detection,kovalev2024detection} to identifying SB2, utilizing all available \lamostmrs\ spectral data, with a primary focus on DR11 and DR12. The data processing followed the same procedure as outlined in \citet{kovalev2024detection}. Example of spectral fitting is shown in Figure~\ref{fig:examplefit}. We converted the heliocentric wavelength scale in the observed spectra from vacuum to air, selected only spectra stacked within a single night, and applied a cut of the signal-to-noise ratio (S/N), retaining only spectra with S/N$\ge 20$ pix$^{-1}$ in any of the spectral arms. More than 12,000 SB2 systems candidates were identified. For orbit fitting, we selected targets with at least six observations, which have changes in radial velocities, resulting in a total of 1,119 sources. For each of these sources, we calculated the radial velocities (RVs) for every observation.
% good now

Before proceeding with the orbital fitting of the radial velocities for SB2 systems, it is crucial to accurately assign the measured radial velocities from a given spectrum to each of the two components. The spectrum fitting methods \citep{kovalev2022detection, kovalev2024detection} generally assign the radial velocities of the more luminous star as \rvone\ and that of the fainter star as \rvtwo. However, this approach is not always reliable and therefore a dedicated radial velocity sorting algorithm is used here to improve accuracy. When \rvone\ and \rvtwo\ are correctly assigned, they exhibit a near-perfect anti-correlation. Plotting \rvone\ against \rvtwo\ should yield points that fall along a straight line, where the slope corresponds to the negative inverse of the mass ratio between the two components. We utilized this method to verify and correct the assignments of \rvone\ and \rvtwo. After the initial processing, where the radial velocity of the more luminous star was assigned to \rvone\ and the fainter star's to \rvtwo, we analyzed the \rvone\ versus \rvtwo\ plot. If any points deviate from the expected line, but swapping \rvone\ and \rvtwo\ for those specific points brings them back onto the line, we re-assign the pairs of radial velocities accordingly. Finally, the data points are correctly attributed to components 1 and 2, and the slope is well defined.

\section{Radial velocities fitting} \label{sec:method}
We used the \python\ package \thejoker\ \citep{price2017joker} for orbit fitting. The original version of \thejoker\ is designed for SB1 and can not fit the \rvone\ and \rvtwo\ simultaneously. The authors of \thejoker\ have provided the SB2 version of the program on \github. To adapt it for our analysis, we made small modifications and enhancements to the SB2 branch, improving its ability to perform orbit fitting both accurately and efficiently.

For SB2 fitting, the radial velocities \rvone\ and \rvtwo\ at time $t$ are given by the following equations:
\begin{align}
\rvone &= v_0 +K_1[\cos(\omega+f) + e\cos\omega] \\
\rvtwo &= v_0 -K_2[\cos(\omega+f) + e\cos\omega]
\end{align}

where $v_0$ represents the systemic radial velocity, and $K_1$ and $K_2$ denote the radial velocity semi-amplitude of star 1 and star 2, respectively. The eccentricity is given by $e$  while $\omega$ represents the argument of periastron. The true anomaly, $f$, is a function of $(t, P, e, M_0)$, where $P$ is the period and $M_0$ is the mean anomaly at the reference epoch. 

To perform the orbit fitting using \thejoker, we assign the following prior distributions to the parameters: the prior for $P$ is set as $\ln P \sim \mathcal{U}(0.3, 1000)~\textrm{days}$, where $\mathcal{U}$ denotes a uniform distribution; the eccentricity $e$ follows a Beta distribution with $e \sim \textrm{Beta}(0.867, 3.03)$. The Beta distribution is a flexible probability distribution defined on the interval $[0, 1]$, commonly used to model proportions and probabilities. The shape parameters of the Beta distribution, $a = 0.867$ and $b = 3.03$, determine the distribution's skewness, with this particular choice favoring smaller eccentricities while allowing for a long tail toward larger values.

Both mean anomaly $M_0$ and the argument of periastron $\omega$ are uniformly distributed between $0$ and $2\pi$. The additional 'jitter' incorporated in quadrature with each visit's velocity error, $\ln s$, follows a normal distribution $\ln s \sim \mathcal{N}(\mu_s, \sigma_{s}^2)$, where $\mu_s$ is the mean of $\ln \epsilon$ in our dataset and $\sigma_s$ is set to 1. The $K_1$, $K_2$ and $v_0$ follow normal distributions as follows,  
\begin{align}
K_1 &\sim \mathcal{N}(0, \sigma_{K}^2)~\kms \\
K_2 &\sim \mathcal{N}(0, \sigma_{K}^2)~\kms \\
v_0 & \sim \mathcal{N}(0, \sigma_{v_0}^2)~\kms
\end{align}
where,
\begin{align}
\sigma^2_K &= \sigma^2_{K, 0}\left(1 - e^2\right)^{-1}\left(\frac{P}{P_0}\right)^{-2/3},\\
\sigma_{K,0} &= 30~\kms,\\
P_0 &= 365~\dayd,\\
\sigma_{v_0} &= 100~\kms.
\end{align}
These prior distributions are similar to those in \stable\ 1 of \citet{price2020close}.

Based on these prior distributions, we sampled 4 million joint samples and then applied rejection sampling. As a result, we obtained the fitting results for $(P,e,\omega,M_0,K_1,K_2,v_0,s)$. However, due to the coupling relationship between $e$ and $\omega$, different combinations of ($e, \omega$) can yield similar radial velocity curves, leading to uncertainties in both $e$ and $\omega$. Therefore, after determining $P$, $K_1$, $K_2$ and $v_0$, we refit $e$, $\omega$, $M_0$ and $s$ to obtain their optimal values. These represent our final orbital parameters. All parameters have errors provided by a following MCMC sampling.

\begin{figure}[ht!]
\centering
\includegraphics[width=0.96\linewidth]{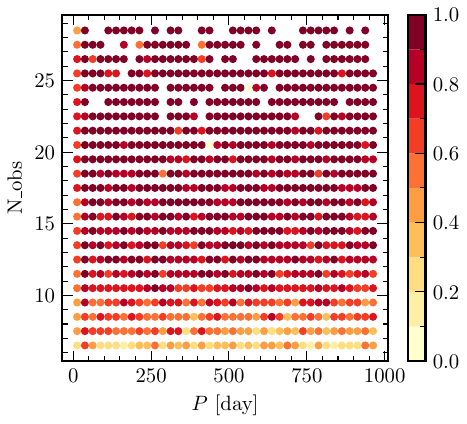} % 20241125_094340, 20241125_221935
\caption{The dependence of the efficiency of orbital fitting on the period $P$ and number of observations $N_\mathrm{obs}$ of SB2s.}
\label{fig:detect}
\end{figure}

To access the efficiency of orbital fitting with \thejoker, we simulated 10,000 SB2 systems with varying orbital periods ($P$), eccentricities ($e$), and mass ratios ($q=M_2/M_1$). The orbital periods $P$ follows a log-uniform distribution ranging from 0.3 to 1,000 days, while the eccentricities $e$ are uniformly distributed between 0 and 0.9, and $q$ are uniformly distributed between 0.1 and 1. The observation epochs and number of observations are consistent with those of the sources in our study. We applied \thejoker\ to fit the simulated orbits generated from these distributions. The fitting efficiency was defined as the fraction of orbits for which the fitted $P$ is within 20\% of the true $P$. The resulting detection efficiency, as a function of $P$ and the number of observations $N_\mathrm{obs} < 10$, is shown in Figure \ref{fig:detect}. 
The efficiency of orbital fitting is generally consistent across different periods, except for cases with $N_\mathrm{obs} < 10$, where shorter-period orbits exhibit slightly higher efficiency. The efficiency remains high (close to 1) for larger $N_\mathrm{obs}$, particularly when $N_\mathrm{obs} > 16$. The trend of the points in Figure \ref{fig:detect} is not entirely consistent, with some outliers. This is because the fitting reliability is influenced not only by $P$ and $N_\mathrm{obs}$ but also by the distribution of observations within the orbit. Given the relatively small sample size of fits (we only simulated 10,000 orbits), such variations are expected. Although the fitting reliability is relatively low for $N_\mathrm{obs} < 10$, these sources still allows for obtaining a reasonably representative set of reliable orbital solutions.

After fitting the RVs, some sources in our analysis yielded multiple similar and nearly equivalent periods in \thejoker\ fitting results, with differences mainly in the eccentricity or the argument of periastron. For these cases, we refined the fitting process around the identified periods. These refined results typically converged to a single orbital solution, where the variations in eccentricity or the argument of periastron are reflected in the uncertainty estimates. For sources where \thejoker\ fitting identified multiple, distinctly different periods, we retained all these potential orbital solutions for further analysis. This approach ensures that plausible orbital configurations are not prematurely excluded.

\section{Fitting Validation and Reliability Constraints}\label{sec:fit_quality}

After obtaining orbital solutions from Section \ref{sec:method}, it is essential to evaluate their reliability and include only robust results in the final sample. To achieve this, we apply a series of validation metrics and constraints to filter out unreliable fits. These metrics assess the goodness of fit and the consistency of the solutions. By refining the results in this way, we aim to establish a high-confidence sample for further analysis.

\subsection{Reduced chi-square}

The reduced chi-square $\chi^2_{\text{red}}$ is a commonly used metric to assess the goodness-of-fit of models to the data and is defined as follows \citep{andrae2010chi}: 
\begin{equation}
    \chi^2_{\rm{red}} = \frac{1}{\nu} \sum_{i=1}^{N} \frac{\left( RV_{\rm{obs}} - RV_{\rm{fit}}\right)^2}{\sigma_i^2}  
    \label{eq:red_chi}
\end{equation}
where the degrees of freedom $\nu$ are calculated as $\nu = 2N-8$, with $N$ being twice the number of observations (since we have both \rvone\ and \rvtwo\ for each observation) and $8$ representing the number of parameters. Here, $\sigma_i^2 = \epsilon_i^2+s^2$ where $\epsilon_i$ is the error in the radial velocities and $s$ is the radial velocity jitter from \thejoker. For a linear model, the expected value of $\chi^2$ for a correctly described dataset is $\chi^2 = \nu$, which implies that $\chi^2_{\text{red}} = 1$. However, given the non-linear nature of the radial velocity models employed, relying solely on $\chi^2_{\text{red}}$ to evaluate the goodness of fit has certain limitations \citep[see][]{andrae2010chi}. 

When the differences between the fitted curve and the data points are comparable to $\sigma$, the $\chi^2_{\text{red}}$ typically approaches unity. However, radial velocity measurements are influenced by various factors, such as stellar pulsations and rotation, which can result in significant deviations for some data points relative to the fitted orbital RV curve. Additionally, the $\epsilon$ in RV measurements are often underestimated, leading to substantially elevated reduced chi-square values for a significant fraction of sources. We simulated data for 800 sources, and the fitting results were categorized by the human eye into well-fitted and poorly-fitted sources. Different $\chi^2_{\text{red}}$ and $NMAE$ values (Section \ref{subsec:nmae}) were selected for data selection, and the accuracy and completeness rates for various filtering criteria were calculated. The results are shown in Figure \ref{fig:scatter_chi_nmae}. It can be seen that using $\chi^2_{\text{red}}$ alone is insufficient to distinguish well-fitted sources.

\subsection{Normalized Mean Absolute Error}\label{subsec:nmae}

\begin{figure*}[ht!]
\centering
\includegraphics[width=\linewidth]{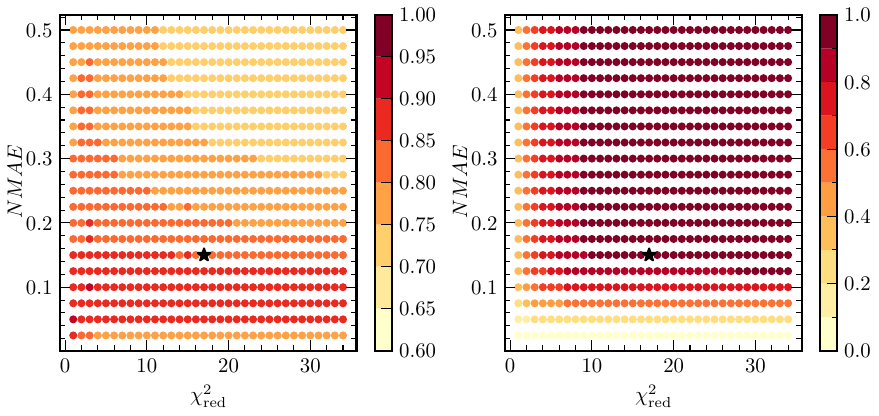}
\caption{Accuracy and completeness rates for different combinations of $\chi^2_{\rm{red}}$ and $NMAE$. Left: Accuracy, defined as the proportion of well-fitted sources among those selected by criteria of $\chi^2_{\rm{red}}$ and $NMAE$ values smaller than the current thresholds. Right: Completeness rate, defined as the proportion of well-fitted sources selected by these criteria relative to all well-fitted sources. The black pentagram represents the final selection criteria: $\chi^2_{\rm{red}} < 17$ and $NMAE < 0.15$.}
\label{fig:scatter_chi_nmae}
\end{figure*}

While the reduced chi-square is not ideal for our case due to the non-linear nature of radial velocity models. To address these challenges, we employed alternative approaches to assess the quality of the fits. Specifically, we use the mean absolute error (MAE), which is less sensitive to outliers compared to the mean squared error (MSE). In periodic fitting, it is important to consider the relative size of the deviations with respect to the amplitude of the radial velocity curve. 
\begin{figure}[ht!]
\centering
\includegraphics[width=\linewidth]{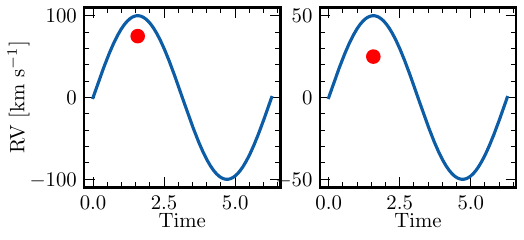}
\caption{Radial velocity curves for two binary systems with identical $P$, $e$, $\omega$, and $t_0$, but different $K$ values. Left: $K = 100$ km/s, Right: $K = 50$ km/s. The two binary systems have same $\chi^2_{\text{red}}$ but solution to the left is better.}
\label{fig:rv_k}
\end{figure}

As illustrated in Figure \ref{fig:rv_k}, we compare two binary systems with identical $P$, $e$, $\omega$, and $t_0$ but different $K$ values: one with $K=100$ \kms and the other with $K=50$ \kms. Both systems have the same observation epochs and number of data points. Except for the points shown in the figure, all other points align with the fitted RV curve. In both cases, the residuals from the fitted curve are 25, meaning the absolute errors are identical. However, due to the different $K$ values, the point in the left plot (with $K=100$ \kms) is closer to the fitted curve, while the fit in the right plot (with $K=50$ \kms) is worse. This demonstrates how the value of $K$ can significantly impact the quality of the fit. To account for this effect, we normalize the mean absolute error (NMAE) by dividing it by the velocity semi-amplitude $K$, thereby reflecting the relative size of the deviations. The $NMAE$ is defined as:

\begin{equation}
NMAE = \frac{1}{2N\,K} \sum_{i=1}^{N} |RV_{\rm{obs}} - RV_{\rm{fit}}|
\label{equ:nmae}
\end{equation}

The metric $NMAE$ can roughly indicate the quality of a fit. Generally, the smaller the $NMAE$, the better the fit. In Figure \ref{fig:scatter_chi_nmae}, we simulated data for 800 sources, and the fitting results were categorized by the human eye into well-fitted and poorly-fitted sources. Different $\chi^2_{\text{red}}$ and $NMAE$ values were selected for data filtering, and the accuracy and completeness rates for various filtering criteria were calculated. It can be seen that using $\chi^2_{\text{red}}$ alone is insufficient to distinguish well-fitted sources and a combination of $\chi^2_{\text{red}}$ and $NMAE$ yields relatively better results. After reviewing the results, we selected $\chi^2_{\text{red}}<17$ and $NMAE<0.15$ as our selection criteria, which give an accuracy of approximately $85$\%  and a completeness of approximately $95$\%.

\subsection{Max phase gap}
In addition to evaluating the quality of the fit, we also assess the orbital phase coverage using the maximum phase gap ($max\_phase\_gap$) as a key metric. This metric represents the largest phase interval without observational data and is used in the validation of radial velocity curve fitting \citep{chen2023new}.

To calculate $max\_phase\_gap$, we first sort all calculated phases in ascending order and compute the phase differences between consecutive sorted phases, taking into account the periodic boundary condition (i.e., including the gap between the smallest and largest phase values). The largest of these differences is then identified as $max\_phase\_gap$. A larger $max\_phase\_gap$ indicates poorer orbital coverage, suggesting that the orbit is poorly constrained in this region. To ensure an adequate orbital solution, we set a threshold where $max\_phase\_gap < 0.35$, meaning that the observations cover at least 65\% of the orbital phase.

\subsection{Radial Velocity Distribution Metrics}

In addition to evaluating phase coverage, we also examine the distribution of radial velocity measurements, which is vital for reliable orbital parameter determination. Uniform sampling across the RV range \citep[e.g.,][]{troup2016companions} and adequate coverage near the extreme regions of the fitted RV curve are essential for constraining parameters like eccentricity ($e$) and the argument of periastron ($\omega$), which can cause significant RV variations over narrow phase intervals. To address this, we introduce a combined metric that evaluates both the uniformity of RV data distribution and the coverage near the curve’s extreme values, ensuring consistent sampling and robust parameter constraints.

To quantify the evenness of data point distribution, the interval $[a, b]$ -where $a =$ rv\_fit\_min and $b = $rv\_fit\_max -is divided into $N_r$ equal sub-intervals. The number of points in each sub-interval ($n_i$) is then compared to the ideal uniform distribution, defined as $n_0 = \text{total points}/N_r$, using the relative deviation:
\begin{equation}
U = \frac{1}{N_r} \sum_{i=1}^{N_r} \left| \frac{n_i - n_0}{n_0} \right|
\end{equation}

\begin{figure*}[ht!]
\centering
\includegraphics[width=\linewidth]{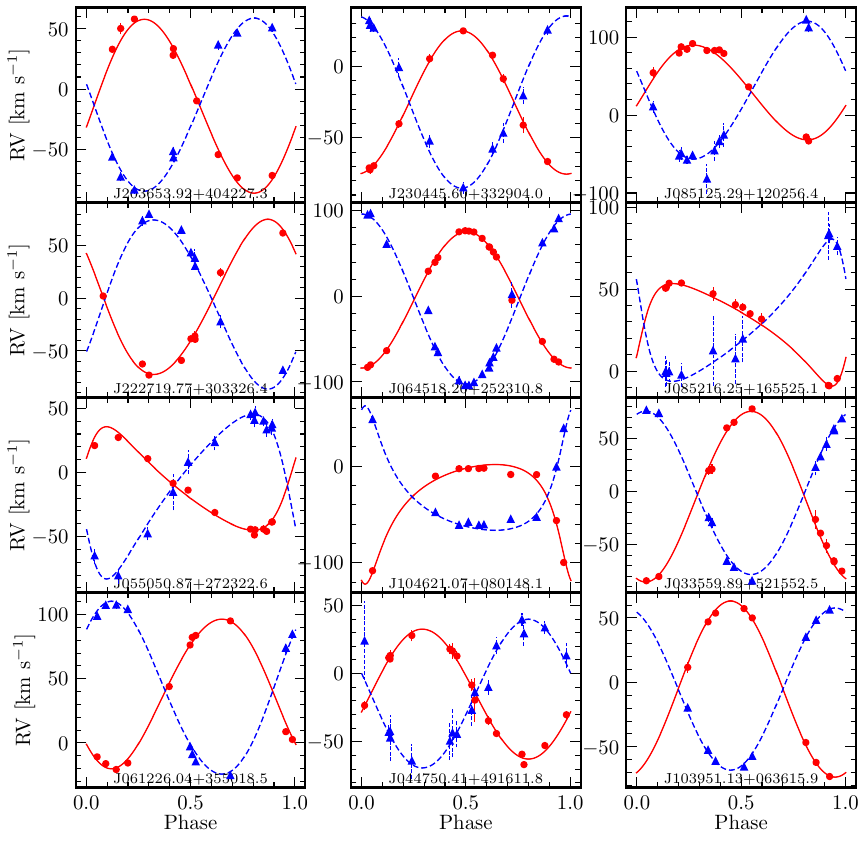}
\caption{Phase-folded radial velocity curves for 12 randomly selected sources. The red solid lines and filled circles represent the fitted curve and measured radial velocity values of the primary star, while the blue dashed lines and triangles represent the fitted curve and measured radial velocity values of the companion star. The radial velocity errors are indicated using error bars.}
\label{fig:example12}
\end{figure*}

To evaluate coverage near the extreme RV regions, we define a simple metric ($B$) that indicates whether these regions are sampled. The extreme RV regions are defined as small intervals around the minimum and maximum fitted RV values, specifically $[a - \epsilon K, a + \epsilon K]$ and $[b - \epsilon K, b + \epsilon K]$, where $\epsilon$ is a parameter (less than 1) determining the size of these regions as a fraction of the velocity amplitude $K$. If at least one data point lies within these intervals, we set $B = 1$; otherwise, $B = 0$. This method ensures that observations near the extreme RV regions are accounted for, even when the total number of observations is small (typically fewer than 20 in our case), without requiring a specific number of points in these regions.

We calculated the $U$ and $B$ values for each component and then averaged them to determine the overall $U$ and $B$ for the binary star system. A smaller $U$ reflects a more uniform distribution, while a larger $B$ indicates better extreme RV regions coverage. The combined metric effectively balances these two aspects to evaluate the quality of the radial velocity data distribution as follows:
\begin{equation}
M = \alpha U + \beta (1 - B)
\end{equation}

Here, $\alpha$ and $\beta$ are weights to determine the relative importance of $U$ and $B$ with $\alpha + \beta=1$. A smaller $M$ indicates better overall performance, with a more uniform distribution and better extreme RV regions coverage. By adjusting $\epsilon$, $\alpha$, and $\beta$, we can fine-tune the metric to align with our requirements. A smaller $\epsilon$ focuses on points closer to the extreme RV regions. A higher $\beta$ emphasizes boundary coverage. For our analysis, we set $\epsilon=0.2$, $\alpha=0.3$, and $\beta=0.7$ to calculate the combined metric $M$. This metric provides a clear quantitative measure that balances uniformity and extreme RV regions coverage, making it easy to compare different datasets or fitting results.

\section{results}\label{sec:results}

\begin{figure*}[htbp]
\centering
\includegraphics[width=0.92\linewidth]{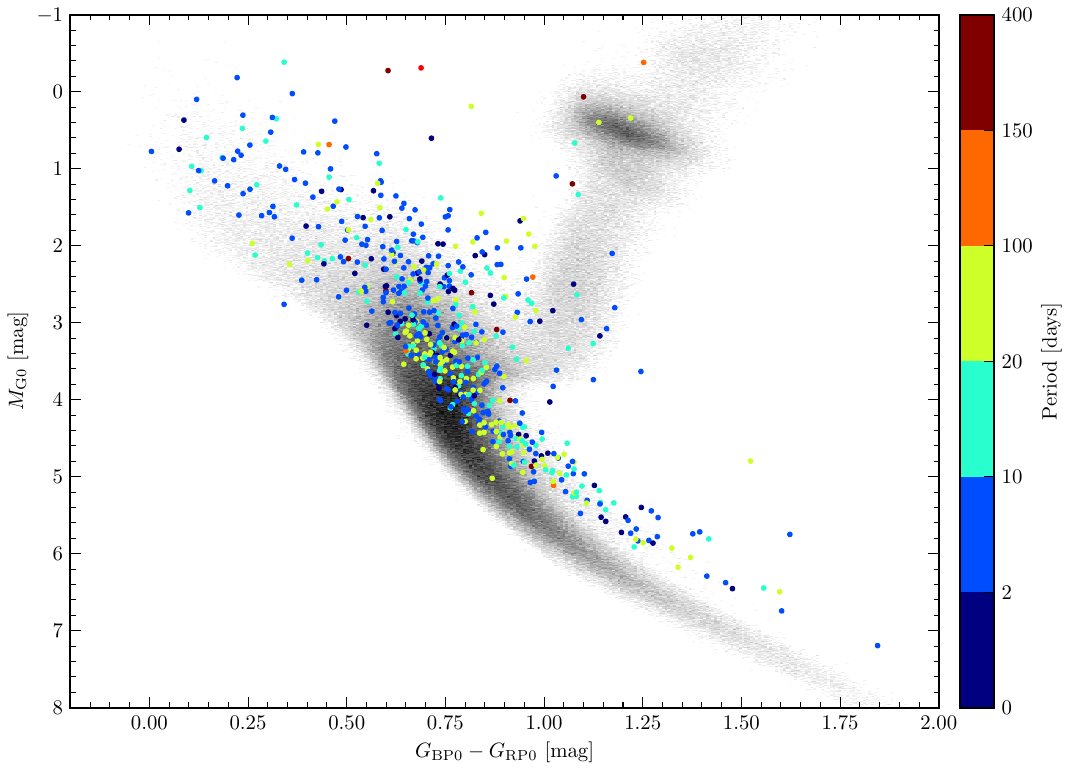}
\caption{Hertzsprung-Russell (HR) diagram comparing the SB2 systems to the full \lamostmrs\ DR11 sample. The SB2s are predominantly located along the binary sequence.}
\label{fig:HR}
\end{figure*}

\begin{figure*}[htbp]
\centering
\includegraphics[width=0.92\linewidth]{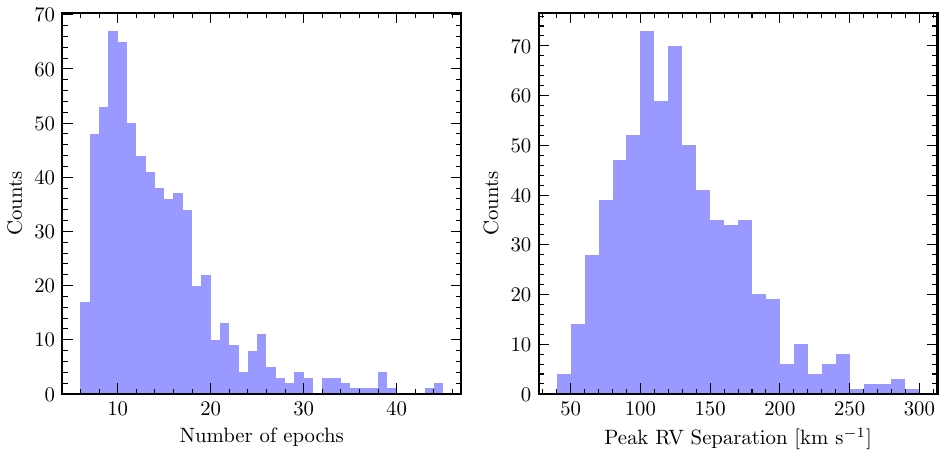}
\caption{Left: Distribution of the number of observational epochs for the SB2s. Right: Distribution of the maximum radial velocity separation between the primary and secondary components.}
\label{fig:hist_epoch}
\end{figure*}

After fitting the 1119 sources with observation times $\geq 6$, we applied three criteria to select reliable orbit fits, as described in Section \ref{sec:fit_quality}:
\begin{enumerate}[topsep=0pt, partopsep=0pt, itemsep=0pt, parsep=0pt]
\item $\chi^2_{\rm{red}}< 17$.
\item $NMAE < 0.15$ to minimize large deviations from the fitted curve.
\item $max\_phase\_gap < 0.35$ to ensure sufficient phase coverage.
\item $M < 0.6$ to account for both the uniformity of the RV data distribution and the coverage of the extreme RV regions.
\end{enumerate}

Using these thresholds, we identified 538 sources with relatively good fits and assigned $fit\_flag = 1$ to them. For some sources with very small $NMAE$ (i.e., $NMAE < 0.06$), we extended the $max\_phase\_gap$ criterion to 0.45. These sources, totaling 56, were also considered to have relatively good fits, and we assigned $fit\_flag = 2$ to them. We conducted a visual inspection of the remaining sources and identified 71 with relatively good fits. Most of these sources have good $NMAE$ values but slightly larger $max\_phase\_gap$ than our criteria. We consider these orbits acceptable so we retained these sources in our final catalog and tagged them with $fit\_flag = 3$. As a result, our final catalog contains 665 sources.

Figure~\ref{fig:example12} presents the orbital fits to the radial velocity curves, showcasing the obtained orbital parameters for 12 randomly selected sources. The figure demonstrates that the fitting results are generally reliable, with the models aligning well with the observed radial velocity data, even in cases with limited observational epochs or moderate uncertainties. Figure \ref{fig:HR} shows the H-R diagram of these sources, which are located in the binary sequence. 

\begin{figure}[htbp]
\centering
\includegraphics[width=0.92\linewidth]{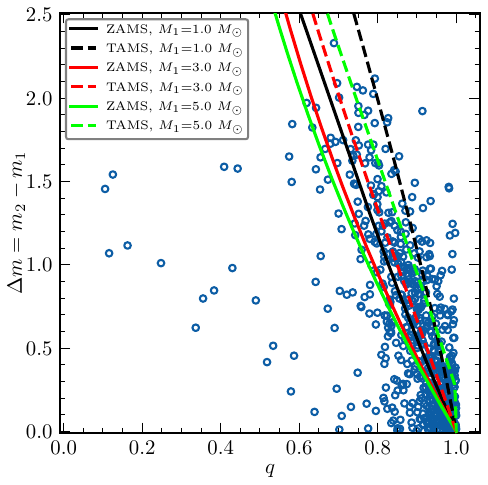} 
\caption{Plot of the magnitude difference ($\Delta m$) versus the mass ratio ($q$). The mass ratio for systems with $q > 1$ is converted to its reciprocal, ensuring that all plotted $q$ values are less than or equal to 1. The $\Delta m$ values are calculated as $\Delta m=|m_2-m_1|$. The black, red and lime lines represent the Zero-Age Main Sequence (ZAMS) of the primary star with $M_1=1.0M_\odot,~3.0M_\odot,~5.0M_\odot$, respectively. The dashed lines correspond to the Terminal-Age Main Sequence (TAMS) of the primary star.}
\label{fig:scatter_detm_q}
\end{figure}

Figure \ref{fig:hist_epoch} presents the histogram of the number of observation epochs and the histogram of the peak RV separation. Most of our SB2s have fewer than 20 observation epochs, and the minimum peak RV separation is 40 \kms, which is the limitation of the method to separate the two components. It is smaller than $50~\kms$ limit in \citet{li2021double}, because the usage of spectral binary model allows to detect SB2s, even from slightly blended spectra \citep{j115}. The maximum peak RV separation is 300 \kms. We measured the magnitude difference between the two components of the binary system. The magnitude difference is derived from the luminosity ratio obtained from the best-fitting solution across multiple observations. Figure~\ref{fig:scatter_detm_q} shows the mass ratio plotted against the magnitude difference. We simulated 10,000 binary systems with different primary masses ($M_1$) using the BSE codes. The black, red, and lime lines represent the Zero-Age Main Sequence (ZAMS) of the primary stars with $M_1 = 1.0M_\odot$, $3.0M_\odot$, and $5.0M_\odot$, respectively, while the dashed lines correspond to the Terminal-Age Main Sequence (TAMS) of the primary stars. Most data points follow a relationship consistent with main-sequence stars. The left outliers are likely evolved systems with a mass transfer history, where the luminosity of the initially more massive component remains relatively unchanged while the other companion’s luminosity increases, causing the points to shift downward.

\begin{figure*}[htbp]
\centering
\includegraphics[width=\linewidth]{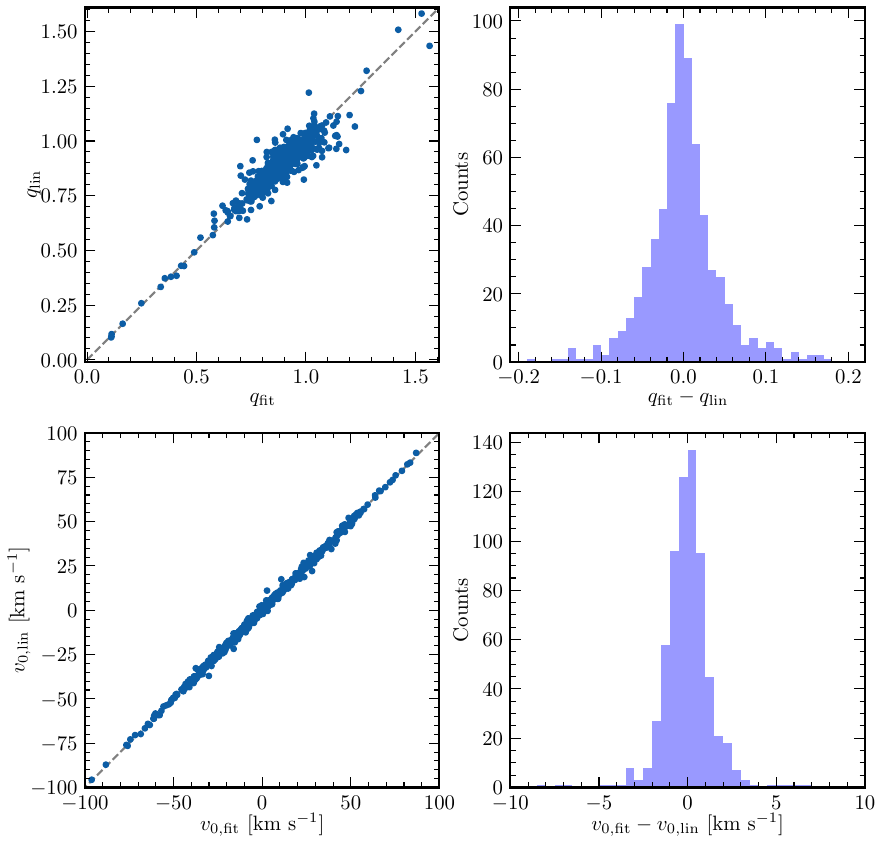}
\caption{Comparison of the inferred mass ratios and $v_0$ for our set of 665 SB2 binary systems. The mass ratio $q_{\text{orb}}$ is calculated as $q = K_1 / K_2$, while $q_{\text{lin}}$ and $v_{0,\text{lin}}$ are obtained from a orthogonal distance regression fit of the radial velocities of the primary and secondary components across multiple epochs, using a Wilson plot. The median absolute difference between $q_{\text{fit}}$ and $q_{\text{lin}}$ is 0.020, while the median absolute difference between $v_{0,\text{fit}}$ and $v_{0,\text{lin}}$ is 0.658.
}
\label{fig:comp_q_v0}
\end{figure*}

In Figure \ref{fig:comp_q_v0}, we present a comparison between our best-fitting values, $q_{\text{orb}}$ and $v_{0,\text{orb}}$, and the results obtained from an orthogonal distance regression (ODR) \citep{odr} fit of the Wilson plot, $q_{\text{lin}}$ and $v_{0,\text{lin}}$. Overall, the agreement between $q_{\text{orb}}$ and $q_{\text{lin}}$ is quite good, with a median absolute difference of $\text{med} |q_{\text{orb}} - q_{\text{lin}}| = 0.020$, and a corresponding middle 68\% range of ($0.005-0.056$). Similarly, the agreement between $v_{0,\text{orb}}$ and $v_{0,\text{lin}}$ is also good, showing a median absolute difference of $\text{med} |v_{0,\text{orb}} - v_{0,\text{lin}}| = 0.658$, with a middle 68\% range of ($0.188-1.521$). These results demonstrate a high level of consistency, providing robust evidence that our fitted values for $K_1$, $K_2$ and $v_0$, are in strong agreement with the dynamic estimates, further validating the reliability of our fitting parameters.

We present all the SB2s along with their orbital parameters and uncertainties in Table \ref{table:665_binary_table}. This includes the designation from \lamostmrs\ DR11, celestial coordinates, source id from \gaia\ Data Release 3, and the $fit\_flag$. The orbital parameters ($P, e, \omega, K_1, K_2, \omega, v_0$) and their corresponding errors are also provided. Additionally, we list the number of observations ($N_{\rm obs}$) and the three metrics used to select reliable orbits: $max\_phase\_gap$, $NMAE$, and $M$.

\setlength{\tabcolsep}{10pt}
\begin{table*}[htbp]
   \caption{List of the parameters of 665 binary candidates.}
     \label{table:665_binary_table}
   \begin{center}
   \begin{tabular}{llllcl}\hline \hline
 Index & Quantity & Column & Format & Units & Description\\
\hline
1 & Designation  & designation   & Char(19) &     ---        & LAMOST designation\\
2 & \gaia\ DR3 source\_id & gaia\_source\_id& Unsigned Longint  &   ---     & source\_id from \gaia\ DR3\\
3 &  R.A.    &  radeg    & Float  &    degrees     & Right ascension in degrees (J2000)\\
4 &  DEC.   &  decdeg    & Float  &    degrees     & Declination in degrees (J2000)\\
5 &  $fit\_flag$ &  fit\_flag    & Int   &    ---         & fit\_flag \\ \hline
6 &  $P$    &  P  & Float   &    days & Period\\
7 &  $P_{\rm err}$ &  P\_err & Float &  days & Typical error of $P$  \\
8 &  $e$     &  e     & Float   &    ---         & Eccentricity \\
9 &  $e_{\rm err}$     &  e\_err     & Float   &    ---         & Typical error of $e$ \\
10 &  $K_1$ &  K1   &  Float & \kms  & RV semi-amplitude of the primary \\
11 &  $K_{\rm 1,err}$ &  K1\_err   &  Float & \kms  & Typical error of $K_1$ \\
12 &  $K_2$ &  K2   &  Float & \kms  & RV semi-amplitude of the secondary \\
13 &  $K_{\rm 2,err}$ &  K2\_err   &  Float & \kms  & Typical error of $K_2$ \\
14 & $v_0$ & v0 & Float & \kms   & Systemic radial volcity\\
15 & $v_{\rm 0,err}$ & v0\_err & Float & \kms   & Typical error of $v_{0}$\\
16 & $\omega$ & omega & Float & radians   & Argument of periastron\\
17 & $\omega_{\rm err}$ & omega\_err & Float & radians   & Typical error of $\omega$\\
18 & $t_0$ & t0 & Float & Julian Day         & Reference epoch for phase = 0 \\
19 & $t_{\rm 0,err}$ & t0\_err  & Float    &  Julian Day & Typical error of $t_0$\\
20 & $s$ & s & Float & \kms      & Radial velocity jitter  \\
21 &  $q$     &  q     & Float   &    ---         & Mass ratio $q=m_2/m_1=K_1/K_2$ \\
22 &  $q_{\rm err}$     &  q\_err     & Float   &    ---         & Typical error of $q$ \\
%23 &  $f(m)$     &  f\_m     & Float   &    ---         & Mass function $f(m)=$ \\
23 & $N_{\rm obs}$ & N\_obs & Int & ---  & Number of observations  \\
24 & $max\_phase\_gap$  & max\_phase\_gap     & Float     & --- &  Maximum phase gap \\
25 & $NMAE$  & NMAE     & Float     & --- &  Normalized mean absolute error \\
26 & $M$  & M     & Float     & --- &  Metric showing the RV distribution \\
27 & $\Delta m$  & mag\_diff     & Float     & --- &  Difference in magnitude\\
28 & $\sigma_{\Delta m
}$  & std\_mag\_diff     & Float     & --- &  Standard derivation of $\Delta m$ \\
\hline\hline\noalign{\smallskip}
  \end{tabular}
    \begin{tablenotes}
  \item[1] This table is available in its entirety in machine-readable form.
  \end{tablenotes}
  \end{center}
\end{table*}

\section{Discussion of initial results} \label{sec:Discussion}

\subsection{Period Analysis}\label{sec:compare_all}

\begin{figure*}[ht!]
\centering
\includegraphics[width=\linewidth]{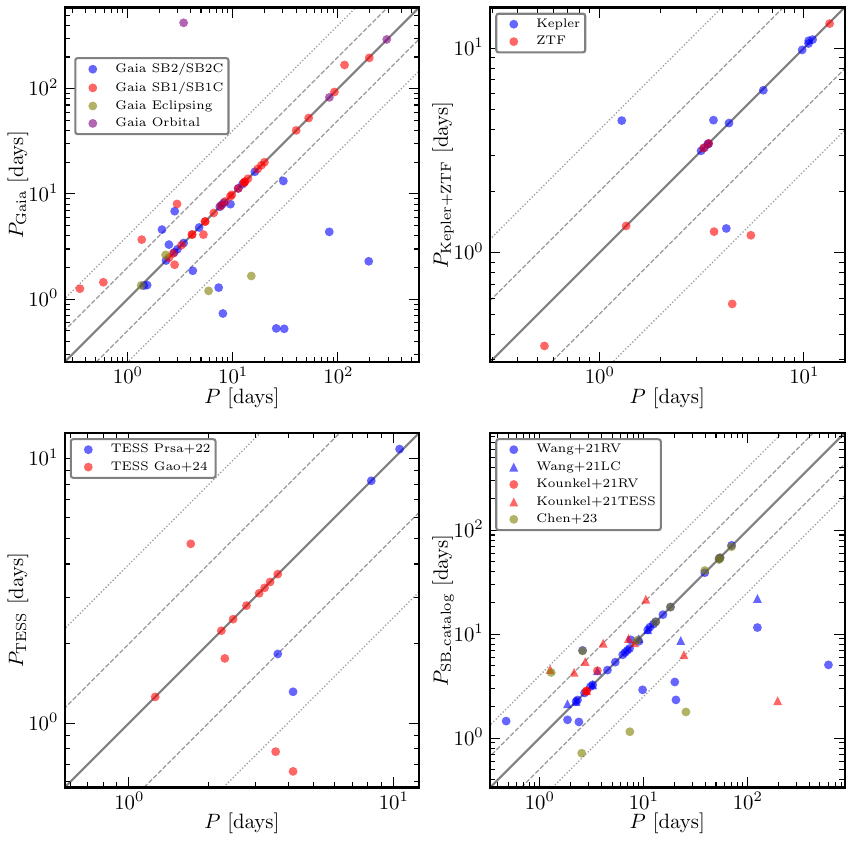}
\caption{Comparison of our periods with those from four variable star catalogs: \gaia\ DR3 non-single stars (top left), Kepler and ZTF (top right), TESS (bottom left), and other SB catalogs from \lamostmrs\ and \apogee (bottom right). The solid grey line represents matching periods, while dashed and dotted lines indicate period differences by factors of two and four, respectively.}
\label{fig:com_p_all}
\end{figure*}

We compared our periods with those from the \gaia\ DR3 non-single star catalog \citep{gaia-nns-2023}, Kepler \citep{prvsa2011kepler1,slawson2011kepler2,matijevivc2012kepler3,conroy2014kepler4,conroy2014kepler5}, ZTF \citep{chen2020ztf}, and TESS eclipsing binaries \citep{prvsa2022tess, ijspeert2021tess} and the periods from radial velocities from \lamostmrs\  \citep{wang2021lamost}, \apogee\  \citep{kounkel2021double} and \lamost-LRS \citep{chen2023new} spectra. The comparisons are illustrated in Figure \ref{fig:com_p_all}. In addition to the solid gray lines indicating periods that match closely, Figure \ref{fig:com_p_all} also includes dashed and dotted grey lines, which represent cases where our periods differ from those in other catalogs by factors of 2 (or 1/2) and 4 (or 1/4), respectively.

In the top-left panel of Figure \ref{fig:com_p_all}, we show the comparison of our periods with those from the \gaia\ DR3 non-single star catalog. There are 63 sources in common, including 23 SB2/SB2C solutions, 31 SB1/SB1C solutions, 4 eclipsing binaries solutions, 3 orbital solutions. 42 of which exhibit matching periods. However we should note that for some sources our periods seems to be more reliable. In particular for J191846.70+433733.4 we found that our period is consistent with the light curve from TESS, while Gaia DR3 value is not.

In the top-right panel of Figure \ref{fig:com_p_all}, we compare our periods with those of variable stars from Kepler and ZTF. All the periods from Kepler correspond to eclipsing binaries, with 13 common sources, 10 of which show very good agreement. There are 8 common sources with ZTF, 4 of which show good agreement.

We found 4 common sources from the eclipsing binary catalog of \citet{prvsa2022tess}, and the periods of 3 agree well. No common sources were found in the catalog of \citet{ijspeert2021tess}. We also searched for common sources in the data from \citet{gao_chen2024tess_variable}, resulting in 12 common sources, after excluding those identified as rotational variables. Among this 12 common sources, 8 show good agreement. These results are presented in the bottom-left panel of Figure \ref{fig:com_p_all}.

In the bottom-right panel of Figure~\ref{fig:com_p_all}, we compare our periods with those obtained from RV fitting using \lamost\ or \apogee\ spectra in the literature. We found 32, 3 and 12 common sources with catalogs derived using radial velocities from \lamostmrs\  \citep{wang2021lamost}, \apogee\  \citep{kounkel2021double} and \lamost-LRS \citep{chen2023new} spectra, respectively, where 32 systems show very good agreement. Additionally, these catalogs provide some photometric periods, where we found 8 common sources from \cite{wang2021lamost} and 10 from \cite{kounkel2021double}. Among these, 13 systems have good agreement with our periods.

In total, we have 63 sources in common with \gaia\ DR3 non-single star catalog, of which most of periods from the SB1 solutions show good agreement. The higher signal-to-noise ratio (S/N) in the LAMOST-MRS spectra makes it possible to detect the faint secondary component in these SB2 systems, whereas the spectra from the \gaia\ Radial Velocity Spectrograph typically reveal only the primary. This contrast stems from LAMOST’s larger mirror size and significantly longer exposure times, compared to \gaia\ \citep{gaia_rvs}. We have 37 common sources with variable star catalogs derived from Kepler, TESS, and ZTF, with 25 showing good agreement. Among these, the agreement of eclipsing binary is particularly strong, with 13 out of 15 matching, reaching an agreement of 86.6\%. We compare our results with the RV fitting results from published catalogs, finding 47 common sources, of which 32 show good agreement. These matching numbers include matches where the periods are half, double, quarter, or quadruple the values in the comparison.

\subsection{Statistics of the catalog}

\begin{figure*}[ht!]
\centering
\includegraphics[width=\linewidth]{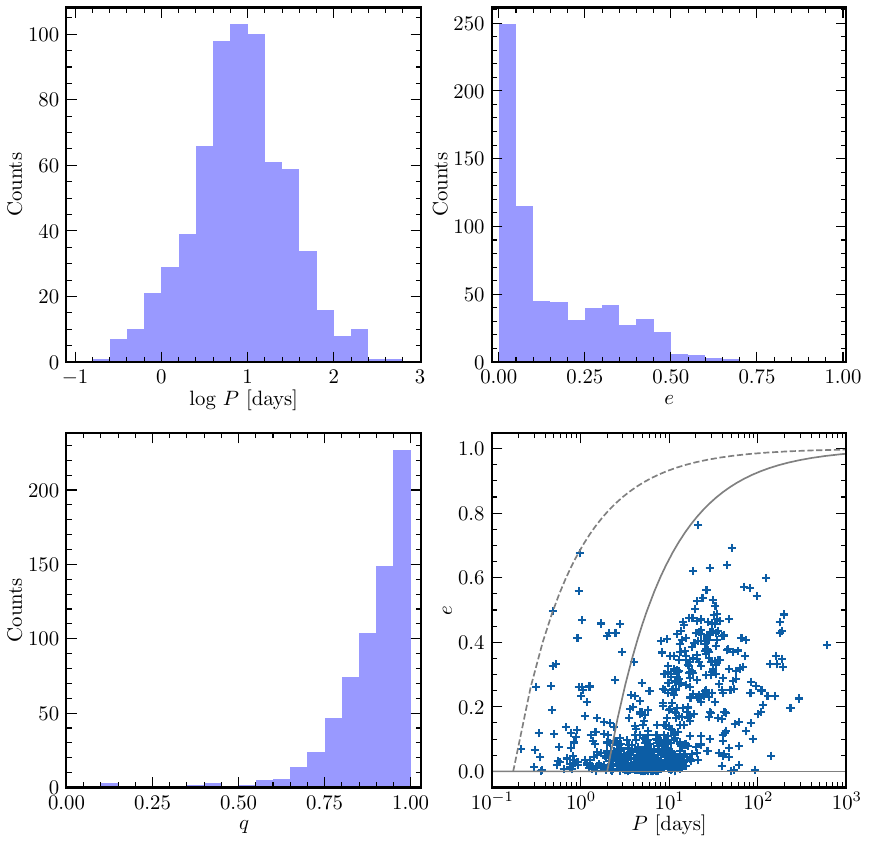}
\caption{ Top left: Histogram of $\log P$. Top right: Histogram of the eccentricity ($e$). Bottom left: Histogram of the mass ratio ($q$), where systems with $q > 1$ are converted to their reciprocal values, ensuring that all plotted $q$ values are less than or equal to 1. Bottom right: Eccentricities as a function of orbital period $P$. The dashed grey curve represents a boundary, beyond which pairs approaching periastron, with a mass sum of 1.5 $M_\odot$, are likely to collide as they pass within 1.5 $R_\odot$ of each other (as adopted by \citet{raghavan2010survey} when investigating the multiplicity of solar-type stars). The solid grey line displays the maximum expected eccentricity as calculated by Equation (3) of \citet{moe2017mind}. }
\label{fig:hist_all}
\end{figure*}

The top-left panel of Figure \ref{fig:hist_all} shows the period histogram, which exhibits a similar log-normal distribution. The top-right panel displays the histogram of eccentricity, revealing that many SB2s in our sample have eccentricities close to 0, indicating nearly circular orbits, while binaries with higher eccentricities are less common. The bottom-right panel shows the period–eccentricity relationship for the 665 SB2s with our final orbital solutions. As expected, most of the shortest-period binaries have circular orbits. Due to dynamical tidal effects that cause orbital circularization, the orbital eccentricities ($e$) of short-period close binaries are expected to be small, generally below 0.2.

The bottom-left panel presents the histogram of the mass ratio. We calculate the mass ratio $q = M_2/M_1 = K_1/K_2$ and convert it to its reciprocal if the original value is greater than 1 to ensure that all plotted $q$ values are less than or equal to 1. The mass ratio of most SB2s is between 0.6 and 1, which is due to the higher selection efficiency of our method for sources with similar luminosity ratios, and hence similar mass ratios. When the luminosity ratio is too small, it becomes difficult to distinguish the spectra of the two stars. This selection effect needs to be accounted for in further statistical analysis (Guo et al in prep).

\subsection{Extreme mass-ratio systems}

\begin{figure*}%[ht!]
\centering
\includegraphics[width=0.3\linewidth]{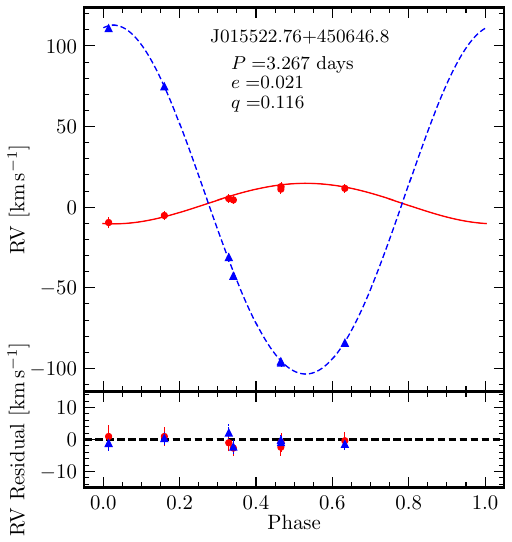}
\includegraphics[width=0.3\linewidth]{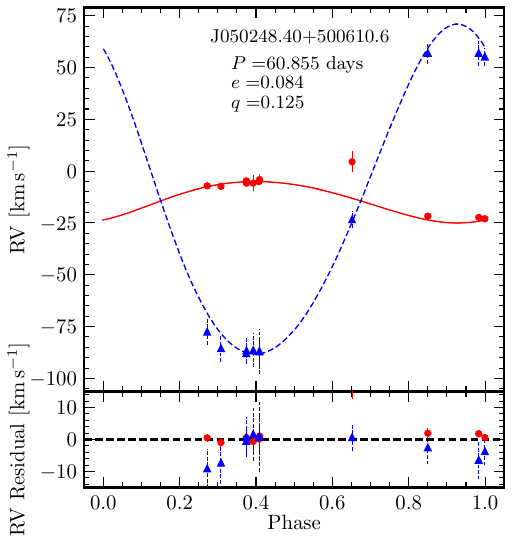}
\includegraphics[width=0.3\linewidth]{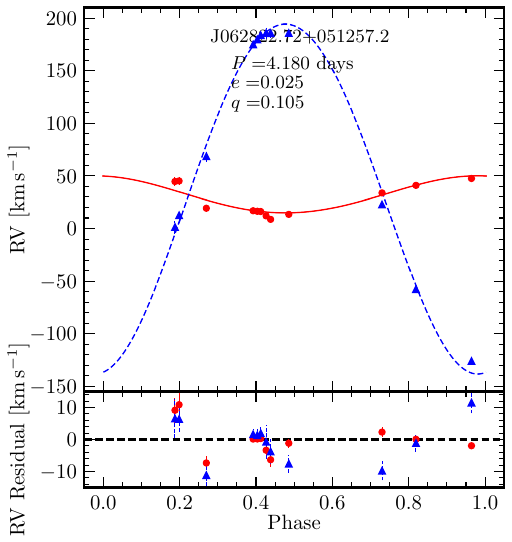}

\caption{Orbital fits for three SB2 systems with the lowest mass ratio in our sample. The red solid line and points represent the fitted curve and actual observed velocity of the primary star, while the blue dashed line and triangles represent the fitted curve and actual observed velocity of the secondary star. The lower panels show the residuals.}
\label{fig:large_q}
\end{figure*}

There are three systems with extremely low mass ratios $q\sim0.1$. Usually for such a big difference of mass the secondary component is almost invisible in the spectra. However once we carefully checked the spectra for these three systems we clearly see both components, moreover for all three of them, the secondary component shows significantly larger Doppler shifts than the primary, see Figure~\ref{fig:large_q}.

Recently, \cite{tvmon} presented a very detailed analysis for one of them: J062822.72+051257.2 aka TV~Mon - a post-mass transfer eclipsing binary of Algol type. They derived a mass ratio $q=0.1056\pm0.0003$, which agrees, within uncertainties, with our estimate.  
Two other systems are also very interesting. The first J015522.76+450646.8 is very similar to the TV~Mon, it has hot primary star $T_{\rm eff}\geq8700$ K contributing to $\sim70$ per cent of the light in the spectrum, while secondary is a cool giant with $T_{\rm eff}\sim5500$ K. The secondary probably fills its Roche lobe, which can be observed as an ellipsoidal variability in the light curve (LC), based on TESS photometry \citep{tess1,tess2}. LC is also consistent with our period value for this system, but shows no eclipses, thus orbital inclination is difficult to constrain. Unfortunately, there are no signs of ongoing mass transfer around the $H_\alpha$ spectral line. We will present more details in our future paper (Kovalev et al in prep.).
Another system J050248.40+500610.6 contains two cool stars with similar $T_{\rm eff}\sim5100$ K that contribute to $75$ and $25$ per cent of the light. Therefore the primary is around two times larger, but almost ten times heavier than the secondary. Their mean density is comparable, so they are likely red giants. The extreme mass ratio and near-zero eccentricity suggest that earlier components were much closer to each other, and there was an intensive mass transfer. The current orbital period $P\sim 60$ days suggests that now the system is wide. TESS data from three sectors show no sign of short-period variability, but also they are unable to confirm the orbital period as the duration of observations in each sector is only 28 days. We plan to analyze this system in details in upcoming paper (Kovalev et al. in prep).

\subsection{Long period SB2 systems}
\begin{figure*}[htbp] 
    \centering  
    \vspace{-0.35cm} 
    \subfigtopskip=2pt 
    \subfigbottomskip=2pt 
    \subfigcapskip=-5pt 
    \subfigure{
        \includegraphics[width=0.45\linewidth]{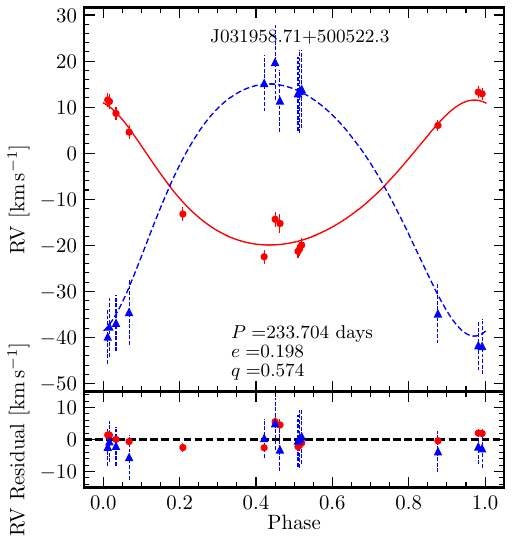}}
    \quad 
    \subfigure{
        \includegraphics[width=0.45\linewidth]{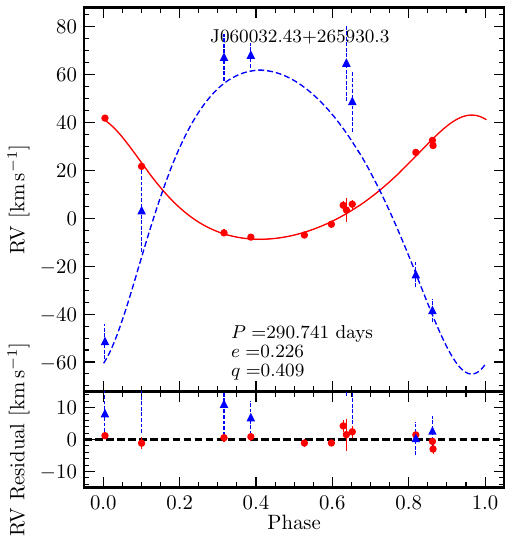}}
    \subfigure{
        \includegraphics[width=0.45\linewidth]{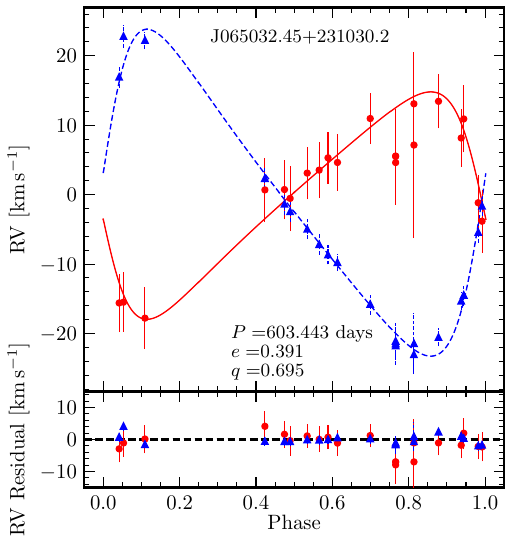}}
    \quad
    \subfigure{
        \includegraphics[width=0.45\linewidth]{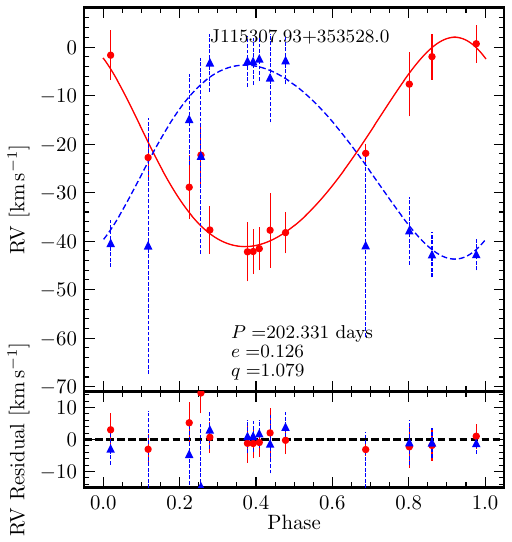}}
    \caption{Orbital fits for four SB2 systems with the period longer than 200 days. The red solid line and points represent the fitted curve and actual observed velocity of the primary star, while the blue dashed line and triangles represent the fitted curve and actual observed velocity of the secondary star. The lower panels show the residuals.}
    \label{fig:longP4}
\end{figure*}

We found several long-period ($P>200$ days) SB2 systems, although only four of them meet our quality criteria. 
\par
J031958.71+500522.3 has an eccentric orbit ($e=0.20\pm0.04$) with a period of $P=233.7\pm0.5$ days. This system has two solutions in \gaia\ DR3: an astrometric orbit ($e=0.23\pm0.06$, $P=240.0\pm0.6$ days, inclination $i=88.89\pm1.44^\circ$) and an SB1 orbit ($e=0.14\pm0.11, P=230.51\pm2.51$ days). These solutions agree well, allowing us to apply Kepler’s Third Law to derive the total mass: 
 \begin{align}
     M_1+M_2=\frac{\left((K_1+K_2)\sqrt{1-e^2})\right)^3}{ GM_{\odot} \sin{i}^3} \frac {P}{2 \pi}=1.80~[M_{\odot}] ,
     \label{eq:kepler31}
\end{align}
 where $GM_\odot=1.32712440041\cdot 10^{20}\, {\rm m^3\,s^{-2}}$  is the Solar mass parameter\footnote{\url{https://iau-a3.gitlab.io/NSFA/NSFA_cbe.html\#GMS2012}}.
Using mass ratio we have $M_1=1.14~M_\odot$ and $M_2=0.66~M_\odot$. Both components are therefore very likely main-sequence stars. We also examined TESS light curves for this system, but found no variability in three sectors.
\par
J060032.43+265930.3 has an eccentric orbit ($e=0.23\pm0.03$) with a period of $P=290.7\pm0.9$ days. The \gaia\ DR3 astrometric solution for this system indicates $e=0.35\pm0.15$, $P=293.7\pm3.9$ days, and inclination $i=84.1\pm1.3^\circ$. The solutions are again consistent, and we use Kepler’s Third Law to find: $M_1+M_2=20.05~M_\odot$, with component masses of $14.24~M_\odot$ and $5.81~M_\odot$. These masses seem to be too large, therefore we suspect this system to be hierarchical multiple system. We also examined the TESS light curve for this system, finding an intriguing result: the light curves for sectors 43, 44, and 45 show eclipses with a period of $P_{\rm ecl}\sim12.35$ days, whereas sector 72 shows no eclipses at all. This unexpected behavior makes it an especially interesting target for future observations, especially using high-resolution spectroscopy.
\par
SB2 J065032.45+231030.2 is an eccentric SB2 system with the longest period in our catalog ($P\sim600$ days, $e\sim0.4$). It comprises a hot, fast-rotating primary and a cooler, slowly rotating secondary. Therefore, spectral components are very different and binary model can easily distinguish them despite blending, making this SB2 the only detection with $K_1+K_2<40$ \kms. Unfortunately, the available public photometry shows no signs of variability for this system. %, which don't allow us to learn about orbital inclination.         
\par
Finally, the twin SB2 system J115307.93+353528.0, was previously studied by \citet{j115} ($e=0.084\pm0.003$, $P=206.41\pm0.08$). It was not originally included in the final version of our catalog because it has NMAE=0.20, but we decided to include it later and set the fit\_flag=3. 
It consists of two nearly identical solar-like stars in a slightly eccentric orbit with a period of $P\sim202$ days and $K_1+K_2\sim40~\kms$. Because of the substantial blending in the \lamostmrs\ spectra, the double-lined structure is difficult to discern. Nevertheless, the binary spectral model successfully recovers radial velocities from these blended lines. Public photometry does not show any obvious variability for this SB2.
\par
It would be highly valuable for future studies to obtain high-resolution follow-up observations of SB2 systems like J065032.45+231030.2 and J115307.93+353528.0. Such observations, for example via high-resolution imaging or interferometric techniques, may resolve these systems optically and enable direct astrometric orbit determinations, thus providing tighter constraints on orbital inclination and a more comprehensive understanding of their fundamental properties.

\subsection{SB1 and another star}
Two SB2 systems J012347.32+410329.1 and J155846.51+443429.4 show no significant changes in RV of the secondary component; thus, for them we can get unrealistic mass ratios. They are similar to the other three systems analyzed in \cite{fakesb2} earlier. Highly likely, they are triple systems where a distant third component is clearly visible in the spectrum, while the inner close system has a very dim or invisible component and therefore can be observed as SB1. There are no other stars in the $5\arcsec$ cone around them, so it is unlikely to be a chance alignment. We analyze these two systems using spectral disentangling similarly to \cite{fakesb2}, but found only two visible spectral components, without detection of the third. Thus we suspect that there is either a white dwarf or a dark star (neutron star or even black hole) in these two systems.

\section{Conclusions} \label{sec:conclusion}

In this work, we have presented the orbital parameters of 665 SB2 binaries from the \lamostmrs\ survey. By employing the identification method of \citet{kovalev2022detection}, we initially selected 1119 SB2s with at least six observations. After fitting their radial velocities using a modified version of \thejoker\ specifically designed for SB2 binaries, we derived the corresponding orbital solutions.

To ensure the robustness and reliability of the obtained orbital parameters, we applied four selection criteria: (1) the reduced chi-square; (2) the normalized mean absolute error, which quantifies the mean deviation of the observations from the fitted RV curve; (3) the $max\_phase\_gap$, which ensures adequate phase coverage; and (4) the RV distribution metric $M$, which checks for a relatively uniform RV distribution with sufficient sampling at the extreme regions of the RV range. After applying these criteria, 665 sources were retained with reliable orbital parameters. In addition to the reduced chi-square, other metrics are also employed to ensure more accurate assessment of the fitting quality. Generally, the more observations there are, the smaller the maximum phase gap, the lower the reduced chi-square, and the smaller the NMAE error, the more reliable the orbital fitting results.

We calculated the magnitude difference between the binary components and compared these values with the corresponding mass ratios. This comparison revealed that most of our data points align with the expected trend for main-sequence stars, indicating a consistent relationship between luminosity and mass ratio. While a few outliers suggest the presence of evolved systems with a mass transfer history, where the initially more massive component retains its luminosity while the companion's luminosity increases.

We compared our results with orbital periods from other well-established catalogs, finding strong consistency with the periods of eclipsing binaries. Specifically, the comparison with data from Kepler, TESS, and ZTF revealed an excellent agreement for eclipsing binaries. The comparison with \gaia\ data yielded good agreement for SB1 solutions from \gaia. These results validate the reliability of our orbital parameter estimations and confirm the consistency of our findings with existing datasets.

The minimum radial velocity separation between the primary and secondary stars is 40 \kms\ in this work, which is constrained by the resolution of  \lamostmrs\ and the nature of our spectral model. Initial statistical analysis of the eccentricity suggests that SB2 systems with periods shorter than 10 days exhibit circular orbits, and the periods follow a log-normal distribution. The statistics of the period, mass ratio, and eccentricity will be analyzed in future work, with careful consideration of selection effects.

%\begin{acknowledgments}
We sincerely thank the referee for their helpful comments and suggestions. Guoshoujing Telescope (the Large Sky Area Multi-Object Fiber Spectroscopic Telescope LAMOST) is a National Major Scientific Project built by the Chinese Academy of Sciences. Funding for the project has been provided by the National Development and Reform Commission. LAMOST is operated and managed by the National Astronomical Observatories, Chinese Academy of Sciences. This work has made use of data from the European Space Agency (ESA) mission Gaia \break (\url{https://www.cosmos.esa.int/gaia}, 
\url{https://gea.esac.esa.int/archive/documentation/GDR3/index.html}).
%\end{acknowledgments}

\vspace{5mm}

\software{Astropy \citep{2013astropyA&A...558A..33A,2018astropyAJ....156..123A,2022astropyApJ...935..167A},
Matplotlib \citep{2007matplotlibCSE.....9...90H},
NumPy \citep{2011numpyCSE....13b..22V},
Joker \citep{price2017joker},
pandas \citep{2010pandas-mckinney-proc-scipy},
SciPy \citep{2020scipy-NatMe..17..261V},
SciencePlots\citep{SciencePlots},
ODR\citep{odr}
}

\bibliography{sample631}{}
\bibliographystyle{aasjournal}

%\end{CJK*}
\end{document}